\relax
\documentstyle[aps,preprint]{revtex}

\newcommand{\la}{\langle}
\newcommand{\ra}{\rangle}
\newcommand{\da}{\dagger}
\newcommand{\p}{\partial}
\newcommand{\be}{\begin{equation}}
\newcommand{\ee}{\end{equation}}
\newcommand{\bea}{\begin{eqnarray}}
\newcommand{\eea}{\end{eqnarray}}
\newcommand{\rd}{\mbox{d}}
\tightenlines
\begin{document}
\bibliographystyle {plain}
\renewcommand{\baselinestretch}{1.26}
\title{Zero Modes and Thermodynamics of Disordered Spin-1/2
Ladders}
\author {Alexander O. Gogolin$^1$,
Alexander A. Nersesyan$^{2,3}$, Alexei M. Tsvelik$^4$ \\
and Lu Yu$^{2,5}$}
\maketitle
\begin {verse}

$^1{\em Imperial~ College,~
Department ~of ~Mathematics,
~180~ Queen`s ~Gate,}$\\
${\em ~London ~SW7 ~2BZ, ~UK,}$\\

$^2{\em International ~Centre ~for ~Theoretical
~Physics,~P.O.Box~586,~34100,~Trieste,}$\\
${\em Italy,}$\\

$^3{\em Institute ~of ~Physics, ~Tamarashvili ~6, ~380077, ~Tbilisi,
~Georgia,}$\\

$^4{\em Department~ of~ Physics,~ University~ of~ Oxford,~ 1~ Keble~ Road,}$\\
${\em~Oxford,~OX1~ 3NP,~ UK}$

$^5 {\em Institute~ of~ Theoretical~ Physics,~ CAS,~  100080~  Beijing,~
China}$\\

\end{verse}
\begin{abstract}
\par

The influence of nonmagnetic doping on the thermodynamic properties
of  two--leg S = 1/2 spin ladders is studied in this paper.
It is shown that, for a weak interchain coupling, the problem
can be mapped onto a model of random mass Dirac (Majorana) fermions.
We investigate in detail the structure of the fermionic states
localized at an individual mass kink (zero modes) in the
framework of a generalized Dirac model.
The low--temperature thermodynamic properties are dominated by
these zero modes. We use the single-fermion density of states,
 known to exhibit the Dyson
singularity in the zero-energy limit, to
construct the thermodynamics of the spin ladder.
In particular, we find that the magnetic
susceptibility $\chi$ diverges at $T\to 0$ as $1/T\ln^2(1/T)$, and
the specific heat behaves as $C\propto 1/\ln^3(1/T)$. The predictions
on magnetic susceptibility are consistent with the most recent results
of quantum Monte Carlo simulations on doped ladders with randomly
 distributed impurities.
We also calculate the average staggered magnetic susceptibility
induced in the system by such defects.
\end{abstract}

\section{Introduction}

Spin ladders have attracted considerable attention of
theorists and experimentalists in recent years \cite{Rice}.
The main distinctive feature of these systems is the
existence of a spin gap in the excitation spectrum
(for ladders with even number of legs).
The spin gap has experimentally been observed, for instance, in
 $SrCu_2O_3$
systems \cite{Srgap}. 
The spectrum of the spin ladders is rather similar
to that of (integer  spin) Haldane chains and spin--Peierls systems.

Somewhat suprisingly, these gapped systems exhibit interesting
behavior when doped by nonmagnetic impurities.
In particular,
$La_{1-x}Sr_xCuO_{2.5}$ compositions have been  investigated
experimentally and a metal-insulator transition was found
\cite{Srdop}. In the low-doping regime ($ x \le 0.05)$ these
systems show an insulating behavior, i.e., holes are localized.
Later on, $Zn$ doping has also been  realized in the
$Sr(Cu_{1-x}Zn_x)_2O_3$ compounds and a transition from the singlet
state to antiferromagnetically (AF) ordered state was observed even
at very low dopings ( of the order $ 1 \%$) \cite{Zndop}.
These doped systems also exhibit a substantial linear in $T$ specific
heat showing abundance of low-energy excitations.
Very recently, the neutron scattering data on these doped
compounds reveal a finite density of states at zero energies,
being consistent with the specific heat data,
 while the amplitude of the spin gap itself remains almost
unchanged\cite{Zudop1}.

Theoretically, the effects of non-magnetic doping in ladder compounds
have been studied extensively, using various numerical techniques
\cite{motome,iino,mdr,mlrd,mikeska,miya}, bosonization\cite{Fuk}, real-space
renormalization group (RSRG)\cite{SF}, nonlinear $\sigma$-model\cite{ng,naga},
Liouville quantum mechanics\cite{Liouville}, the Berezinskii diagram
technique\cite{Berezinskii}, the supersymmetric method
\cite{supersymmetry},  etc. Intuitively,
one might  expect that, because of the spin gap, the
impurities would  be irrelevant and have no significant
effect. However, this is not true. The  nonmagnetic impurities create
 low--energy (in--gap) localized states which dominate the low--temperature
thermodynamics.  Up to now, there is already some consensus in the
theoretical understanding of this issue (without making explicit
references): The nonmagnetic impurity
induces a spin 1/2 degree of freedom around it which leads to a Curie-like
uniform
susceptibility; the effective interaction
between these "free" spins can be ferromagnetic or AF,
depending on the impurity configurations; there  should be zero-energy
states which show up in
neutron scattering experiments and give rise to  additional specific heat;
the AF-magnetic
correlations are enhanced around the impurity, etc.

 However, there are still several
important open questions: What is the exact form of the Griffiths singularity
for the susceptibility (for which the RSRG analysis
\cite{SF} and quantum Monte Carlo (QMC) simulations\cite{miya} gave very
different results)?
How to derive the "zero modes" assumed in several calculations? How  to
calculate the staggered susceptibility? How to construct a self-consistent
theory
for the singlet-AF phase transition?

 In this paper we present a more systematic theoretical study of the
doping effect in two-leg spin 1/2 ladders, based on symmetry analysis,
bosonization technique and mapping on  random mass Dirac(Majorana)
fermion model. We will
limit ourselves to the insulating regime when doped holes are localized.
We first introduce the bosonization technique and
 show explicitly how the presence of holes will modify the motion
of the spin degrees of freedom (first for a single spin 1/2 chain in Section
\ref{imp}). Then in Section \ref{DM} we map the doped
spin ladder system onto a model of Majorana fermions with  random masses.
Moreover, in Section \ref{ZM} we investigate the symmetries
and the states of the fermionic model with a
single mass kink to explicitly derive the zero modes.
In Section \ref{therm} the thermodynamic functions are evaluated
and the calculated unform susceptibility is compared with the
QMC results\cite{miya}. Furthermore, in Section \ref{SS} we calculate the
average
staggered magnetic susceptibility caused by a defect. Finally, some
concluding remarks
are given in Section \ref{CD}.

Some of the presented results are already known to some extent by now,
especially
in view of the similarity between the ladders and the spin-Peierls
systems\cite{fuk1,FM}(although there are some essential differences
between these two cases).  However, we believe our study sheds new light
on the problem and provides justification for several assumptions
made in earlier papers. The comparison with QMC results shows first evidence
of nontrivial physics in the problem--difference between "typical"
and "average" behavior of the spin correlations in systems with
randomly distributed impurities, as will be explained
later in Section \ref{therm}.

\section{Nonmagnetic impurities in itinerant spin--1/2 chains}
\label{imp}

Let $H_{bulk}$ be a standard Hamiltonian for a one--dimensional
interacting electron system.
We shall not write $H_{bulk}$ explicitly in terms of the electron
field operators (its bosonized version is given below),
referring the reader to the literature instead \cite{SE}.
We assume that interaction is spin-rotationally invariant, and
choose the interaction constants in such a way that the spin sector
of the model remains gapless, while charge excitations have a finite gap,
a repulsive half--filled Hubbard model being a typical example.
Although we will only be dealing with the spin degrees of freedom in this
paper,
it is very helpful to include explicitly the charge degrees of freedom as
will be clear from the later presentation.
Then, at energies well below the charge gap,
$H_{bulk}$
describes an itinerant
SU(2)-symmetric spin--1/2 chain.
The electron field operator is bosonized as
\[
\frac{1}{\sqrt{a_0}} \psi_\sigma (x) \rightarrow  e^{ik_Fx}R_\sigma(x)+
e^{-ik_Fx}L_\sigma (x)
\]
with
\[
R_\sigma(x)\rightarrow \frac{1}{\sqrt{2\pi a_0}}
e^{i\sqrt{4\pi}\phi_{R\sigma}(x)}\;,\;\;\;\;\;
L_\sigma(x)\rightarrow \frac{1}{\sqrt{2\pi a_0}}
e^{-i\sqrt{4\pi}\phi_{L\sigma}(x)}\;,
\]
$a_0$ being the short--distance cutoff.
The chiral Bose fields $\phi_{R(L)\sigma}$ combine
in the standard way
to produce the charge phase field $\Phi_c$ and the
spin phase field $\Phi_s$,
as well as the corresponding dual fields $\Theta_{c(s)}$.
As usual, at low energies, the charge and spin degrees of freedom decouple
in the bulk: $H_{bulk}=H_c+H_s$.
The charge Hamiltonian has the form of a quantum sine-Gordon model
\be
H_c=H_0\left[\Phi_c\right]-\frac{m_c}{\pi a_0}\int \rd x
\cos\left[\sqrt{8\pi K_c}\Phi_c(x)\right]\;,
\label{IHc}
\ee
where $m_c>0$ is the
bare
charge mass,
and the phase field is rescaled according to
$\Phi_c\rightarrow \sqrt{K_c}\Phi_c$, with $K_c\leq 1$ being the
charge exponent. Here $H_0$
is the canonical Hamiltonian for the Gaussian model
\[
H_0[\Phi_c]=\frac{v_c}{2}\int \rd x \left[
\Pi_c^2(x)+(\partial_x\Phi_c(x))^2\right]\;,
\]
where $\Pi_c$
is the momentum canonically conjugate to
$\Phi_c$, and $v_c$
is the (charge) velocity.
Up to a marginally irrelevant perturbation,
the Hamiltonian for the spin degrees of freedom is simply
\be
H_s=H_0[\Phi_s]
\label{IHs}
\ee
with the appropriate spin velocity $v_s$.

In the rest of this Section we discuss the effect of
impurities in a single chain.
A single nonmagnetic impurity in spin--1/2 chains has been
studied, for localized spins, by Eggert and Affleck (EA)
\cite{EA}. Let us start by summarizing their results.
EA discovered that there are impurities of two types:

{\bf (L)} Impurities which
violate the site parity $P_S$. These impurities may or may not respect
the link symmetry $P_L$.
An example is an exchange constant modified on a single link.
Such impurities are relevant and break the chain up.
The
infrared stable fixed point corresponds to
an open boundary condition.

{\bf (S)} Impurities which respect $P_S$ (hence violate $P_L$).
An example is two neighbouring exchange couplings modified by the
same amount. These impurities are irrelevant, so that at
low energiers the chain ``heals".

The physical interpretation of EA's findings is as follows.
Given that the spin dimerization is the leading instability of the
Heisenberg spin--1/2 chain, one can immediately see that the
dimerization
order parameter can be locally pinned by the
{\bf L}-type impurities
but
not
by the
{\bf S}-type impurities.
Another way around is to say that the dimerization
operator $\epsilon (x)$
is invariant under $P_L$ but changes its sign under $P_S$:
a local relevant perturbation $\epsilon (0)$ is allowed
for {\bf L}-type impurities, but it is, by symmetry, prohibited for
{\bf S}-type impurities, with $\partial_x \epsilon (0)$ being the
leading irrelevant operator.

In addition to EA's considerations we need to trace the spatial
behavior of the charge phase field $\Phi_c (x)$ in order to
understand how the coupling between the chains is modified by
the presence of these irrelevant impurities. (We imply {\bf S}-type impurities:
see the discussion at the end of the Section.)

Let us first consider the case when the charge sector is gapless.
The system admits an
arbitrary electron charge induced by the impurity potential:
\be
Q=e\sqrt{\frac{2K_c}{\pi}}\int_{-\infty}^\infty \rd x \partial_x
\langle\Phi_c(x)\rangle\;.
\label{IHQ}
\ee
For an impurity localized at the origin over a scale $\sim a_0$
it means that
\be
\langle \Phi_c(x) \rangle = \sqrt{\frac{\pi}{2K_c}}\frac{Q}{e}
\theta (x)\;,
\label{IHphiav}
\ee
$\theta (x)$ being the step function defined by
$\theta(x<0)=0$, $\theta(0)=1/2$, and $\theta(x>0)=1$
[in the gapless case, an arbitrary constant can be added to
Eq.(\ref{IHphiav})].
Eqs. (\ref{IHQ}) and (\ref{IHphiav}) describe the well-known charge
fractionization in a Luttinger liquid \cite{glaz}.

On the other hand, when the charge sector is gapful
(the case we are really interested in),
the ground state expectation value of the charge phase field should coincide
with one of the $Z_{\infty}$-degenerate vacua of the potential in the
sine-Gordon model (\ref{IHc}):
\be
\langle \Phi_c(x) \rangle \rightarrow \sqrt{\frac{\pi}{2K_c}} n
\;\;\;\;\;{\rm as}\;\;\;\;\;
x\rightarrow\pm\infty\;,
\label{IHphicond}
\ee
$n$ being an integer. It is important to notice that no local
impurity potential can overcome the bulk energy of the system.
Hence the asymptotic condition (\ref{IHphicond}) must be
satisfied for any impurity scattering operator.
The integer $n$, however, can vary and may be different for
$x\rightarrow -\infty$ and $x\rightarrow +\infty$; this only
involves a local alternation of the umklapp scattering term.

The condition (\ref{IHphicond}) has an immediate effect on
the value of the electron charge (\ref{IHQ}) that can be trapped by
impurities in a gapped system. Indeed, comparing (\ref{IHphiav})
and (\ref{IHphicond}) one finds that the charge is quantized
\be
Q=em,
\label{IHQint}
\ee
with $m$ being another integer number equal to the increase of $n$
in going from $x=-\infty$ to $x=+\infty$.

For a single impurity it is
natural to
assume that $m=1$ ($m=-1$), so that one electron (hole) is
trapped around the
impurity site\cite{fuk2}.
If many such impurities are scattered along
the system, then the charge phase field acquires an average
value
\be
\langle \Phi_c(x) \rangle=\sqrt{\frac{\pi}{2K_c}}N(x)\;,
\;\;\;\;N(x)=\sum_i\theta(x-x_i)\;.
\label{IHphiavm}
\ee
The bosonized form of the staggered magnetization is then given by \cite{SNT}
\be
n^z(x) =  - \frac{\lambda(x)}{\pi a_0}
\sin \left[\sqrt{2 \pi} \Phi_s(x)\right]
\;,\;\;\;
n^\pm(x) =  \frac{\lambda(x)}{\pi a_0}  \exp\left[\pm i
\sqrt{2\pi} \Theta_s(x)\right]\;,
\label{IHn}
\ee
with the function $\lambda(x)$ defined as
\be
\lambda(x)=\langle \cos\left[\sqrt{2\pi K_c}
\Phi_c(x)\right]\rangle\;.
\label{IHlambdapre}
\ee
Using (\ref{IHphiavm}), one obtains
\be
\lambda (x)=\lambda_0
\exp\left[i\pi N(x)\right]=\lambda_0\prod_i {\rm sign}(x-x_i)\;,
\label{IHlambda}
\ee
where $\lambda_0$ is a non--universal dimensionless constant
equal to
the average (\ref{IHlambdapre}) in the absence of
the disorder.

Similarly,
the spin dimerization
operator acquires an identical $x$--dependent prefactor due to the
alternation of the average value of the charge phase field by the
nonmagnetic impurities:
\be
\epsilon (x) = (-1)^n {\bf S}_n \cdot {\bf S}_{n+1} \rightarrow
\frac{\lambda(x)}{\pi a_0}
\cos\left[\sqrt{2\pi}\Phi_s(x)\right]\;.
\label{IHrho}
\ee
Thus, the only but important effect of {\bf S}-type impurities on the single
chain is
the appearance of the sign alternating factor in the definitions of the
staggered component of the spin density and dimerization field.
The consequences of this factor for the model of coupled chains are discussed
in the next Section.

Here we would like to note that $Sr$ doping of Ref. \cite{Srdop} probably
leads to impurities of the type {\bf S}.
Indeed, $Sr$ doping produces holes in the system.
Given the similarity of the chemical composition of $La-Cu-O$ chain systems
and $LaCu_2O_4$ high-$T_c$ compounds, it is natural to assume that holes
are localized on the oxygen ions (for they are known to be localized
on the oxygen ions in the high-$T_c$ materials).
The localized hole with  spin $1/2$ represents then a ``new"
site in the magnetic chain.
Since this effectively adds an extra site as compared to the pure chain,
the physical meaning of the staggered magnetization sign change
becomes almost trivial.
Since {\bf S}-type impurities are irrelevant in the renormalization
group sense (i.e., the coupling of the hole spin to its neighbours flows
toward the uniform exchange $J$),
this change of the sign is the only effect of $Sr$ doping.

On the contrary, the impurities of {\bf L} type
(like $Zn$ doping of Ref.\cite{Zndop}) are relevant,
so that one may conclude that the chain segment model
must be used.
We would like to point out that the last conclusion
is not always correct.
It is true that {\bf L}--type impurities are relevant with respect
to the single chain ground state, but they are still
irrelevant with respect to the ladder ground state since
the latter has a gap.
Thus, for low concentrations, the {\bf L}-type impurities should
have the same effect as {\bf S}-type impurities, the crossover
 to the chain segment model occuring only at higher
concentrations.
(The crossover concentration is, of course,
exponentially small in the ratio of $J$ to the spin gap.)
Therefore, generically speaking, one should not consider  severed
chains as a  realistic model for nonmagnetic dopings in spin ladders.

\section{Dirac and Majorana fermions with random mass}
\label{DM}

Let us now consider a model of two weakly coupled S = 1/2
Heisenberg chains -- a two--leg spin ladder. Its Hamiltonian,
\be
H=H_0+H_\perp\;,
\label{TDMHtot}
\ee
consists of two terms. The first term,
\[
H_0=\sum\limits_{j=1,2} H_0[\Phi_j]\;,
\]
describes two decoupled chains
(the subscript {\it s} for the `spin' is suppressed since
we shall only work with the spin phase fields in what follows).
The second term
\be
H_\perp=a_0^{-1}\int \rd x \left[ J_\perp\vec{n}_1(x)\vec{n}_2(x)+
U\epsilon_{1}(x)\epsilon_{2}(x)\right]\;,
\label{TDMHperp}
\ee
is responsible for the interchain interaction. Note that only the
relevant interactions are retained while all the marginal terms,
leading to renormalization of the masses and velocities,
are neglected. $J_\perp$ is the interchain exchange coupling.
The second term in (\ref{TDMHperp}), which couples
dimerization order parameters of different chains,
can either be
effectively mediated by spin-phonon interaction \cite{NT} or, in the doped
phase,
generated by the conventional Coulomb repulsion
between the holes moving in the spin correlated background
\cite{Shelton2}.

Using the bosonization formulas (\ref{IHn}) and (\ref{IHrho})
and introducing the symmetric and antisymmetric combinations of the spin
fields,
\[
\Phi_\pm=\frac{1}{\sqrt{2}}\left[
\Phi_1\pm\Phi_2\right]\;,\;\;\;\;\;
\Theta_\pm=\frac{1}{\sqrt{2}}\left[
\Theta_1\pm\Theta_2\right]
\]
one easily finds that, like in the case of
a pure spin ladder \cite{Schulz}, the total Hamiltonian
(\ref{TDMHtot}) factorizes into two commuting parts
\[
H=H_++H_-\;,
\]
with $H_\pm$ given by
\be
H_+=H_0[\Phi_+]+\frac{u-1}{\pi a_0}\int\rd x~
m(x) \cos \left[ \sqrt{4\pi} \Phi_+(x)\right]
\label{TDMHpl}
\ee
and
\bea
H_-=H_0[\Phi_-]&+&\int\rd x \left\{ \frac{u+1}{\pi a_0}
m(x) \cos \left[ \sqrt{4\pi} \Phi_-(x)\right]\right.
\nonumber\\
&+&
\frac{2m(x)}{\pi a_0} \cos\left. \left[ \sqrt{4\pi}
\Theta_-(x)\right]
\right\}\;.
\label{TDMHmin}
\eea
Here $u=U/J_\perp$ and
\be
m(x)=mt(x),~~~
m=\frac{J_\perp\lambda_0^2}{2\pi}, ~~~
t(x)=\exp\left[i\sum\limits_{j=1,2}N_j(x)\right].
\label{TDMtel}
\ee

The function $t(x)=\pm 1$ changes its sign whenever
one passes through a
position of an impurity on either chain.
If one assumes, as we shall do, that the positions of the impurity centers are
uncorrelated, then
the function $t(x)$ represents a random `telegraph process'
characterized by the average density $n_0$ of the impurities,
$a=1/n_0$ being the mean distance between them.
Thus the disorder manifests itself in multiplying the interaction
term by the telegraph signal factor $t(x)$.
This leads to a random mass fermion problem, as we shall see shortly.

Since the scaling dimension of the interaction terms in the
Hamiltonians $H_\pm$ is equal to 1, these can conveniently
be re-fermionized as in the pure case \cite{SNT}.

Let us start with $H_+$.
The chiral components of $\Phi_+$ can be
related to chiral component of a new Fermi field
$\psi$ by the standard bosonization formula
\be
\exp\left[\pm i\sqrt{4\pi}\phi_{+, R(L)}(x)\right]
=\sqrt{2\pi a_0}~ \psi_{R(L)}(x)\;.
\label{TDMpsi}
\ee
In terms of this new Fermi field, $H_+$ becomes
\bea
H_+&=&H_D^{m_t}[\psi]=
\int\rd x \left\{
-i v_s \left[\psi^\dagger_R(x)\partial_x
\psi^{\phantom{\dagger}}_R(x)-
\psi^\dagger_L(x)\partial_x\psi^{\phantom{\dagger}}_L(x)
\right]\right. \nonumber \\&-& im_t t(x) \left.\left[
\psi^\dagger_R(x)\psi^{\phantom{\dagger}}_L(x)-
\psi^\dagger_L(x)\psi^{\phantom{\dagger}}_R(x)]\right]\right\}\;,
\label{TDMHDir}
\eea
where $m_t=(1-u)m$.
Thus, the Hamiltonian $H_+$ describes Dirac fermions with
a random (telegraph signal) mass.
It is sometimes convenient to separate the real (Majorana)
components
of the Fermi field operator
\[
\psi_{R(L)}(x)=\frac{1}{\sqrt{2}}\left[
\zeta^1_{R(L)}(x)+i\zeta^2_{R(L)}(x)\right]
\]
so that
\be
H_D^{m_t}[\psi]=\sum\limits_{a=1,2}H_M^{m_t}[\zeta^a]
\label{TDMDir-Maj}
\ee
with
\bea
H_M^m[\zeta]=&-&\int\rd x \left\{\frac{iv_s}{2}\left[
\zeta_R(x)\partial_x\zeta_R(x)-
\zeta_L(x)\partial_x\zeta_L(x)\right]
\right.\nonumber\\
&+&\left.imt(x)\xi_R(x)
\zeta_L(x)\right\}
\label{TDMMaj}
\eea
standing for the Hamiltonian of the random mass Majorana field.

The Hamiltonian $H_-$ admits a similar re-fermionization
procedure based,
analogously to (\ref{TDMpsi}),
on the introduction of the Fermi field
\[
\exp\left[\pm i\sqrt{4\pi}\phi_{, R(L)}(x)\right]
=\sqrt{2\pi a_0} \chi_{R(L)}(x)\;.
\]
The only difference with (\ref{TDMHDir}) is the appearance of a
`Cooper--type'
mass due to the presence of the dual field ($\Theta_-$) in the
interacting part of $H_-$, Eq(\ref{TDMHmin}).
This can easily be diagonalized by directly passing to the
Majorana components
\[
\chi_{R(L)}(x)=\frac{1}{\sqrt{2}}\left[
\zeta^3_{R(L)}(x)+i\zeta^0_{R(L)}(x)\right]\;.
\]
As a result the Hamitonian $H_-$ decomposes
into two random mass Majorana models
\[
H_-=H_M^{m_t}[\zeta^3]+H_M^{m_s}[\zeta^0]
\]
with different amplitudes of the mass telegraph signal,
$m_t$ and $m_s=-(3+u)m$.

The total Hamiltonian can now be represented in the form
\be
H=H_++H_-=\sum\limits_{a=0}^3H_M^{m_a}[\zeta_a]
\label{IHHtotM}
\ee
($m_0=m_s$, $m_{1,2,3}=m_t$) which,
as in the pure case \cite{Ts1990,SNT},
reflects the spin rotational symmetry of the problem
(SU(2) symmetry is, of course, preserved by the disorder).
The Majorana fields $\zeta^a$ with $a=1,2,3$ correspond
to triplet magnetic excitations, while $\zeta^0$
describes a singlet excitation. All these fields
acquire random masses due to the presence of disorder.
(One should bear in mind that, when marginal interactions are included,
the masses $m_t$ and $m_s$ get renormalized, and
the velocities in the triplet and singlet sectors
become different: $v_s\rightarrow v_t$, $v_s$.)

The mass kinks create low--energy states within the spin gap.
One therefore expects the low--temperature thermodynamic
functions to be dominated by these states.
Before constructing the thermodynamics of the system
(Section \ref{therm}),
we investigate in detail the theory with
an isolated kink.

\section{Zero modes and fermion number in a generalized Dirac model}
\label{ZM}

Let us consider a Dirac Hamiltonian $H$ with the structure of $H_-$ in Eq.
(\ref{TDMHmin}),
assuming that both the Dirac (CDW-like) and Cooper mass functions,
$m_1 (x)$ and $m_2 (x)$,
are of the "telegraph process" type but otherwise are independent.
It is convenient to make a chiral rotation
\begin{equation}
R \rightarrow \frac{R + L}{\sqrt{2}}, ~~~L \rightarrow \frac{- R + L}{\sqrt{2}}
\label{chir.rot}
\end{equation}
under which only the kinetic energy term in (\ref{TDMHmin}) is modified
($\gamma_5 = \sigma^3 \rightarrow \sigma^1$):
\begin{eqnarray}
H (x) = - i v ( R^{\da} \p_x L + L^{\da} \p_x R )
&+& i m_{1}(x) ( R^{\da} L - L^{\da}  R ) \nonumber\\
&+& i m_{2}(x) ( R^{\da} L^{\da} - L R ). \label{Ham.gen1}
\end{eqnarray}
We know that this generalized Dirac Hamiltonian
factorizes into two decoupled massive Majorana fields:
\begin{eqnarray}
R &=& \frac{\xi^+ _R + i \xi^- _R}{\sqrt{2}},
~~~ L = \frac{\xi^+ _L + i \xi^- _L}{\sqrt{2}}, \nonumber\\
H &=& H_{m_+} [\xi^+] + H_{m_-} [\xi^-], \nonumber\\
H_{m_{\sigma}} (x) &=& - i v  \xi^s _R \p_x \xi^s _L
+ i m_{\sigma} (x) \xi^s _R  \xi^s _L, ~~~( \sigma = \pm). \label{Majorana}
\end{eqnarray}
where
$$
m_{\pm}(x) = m_1 (x) \pm m_2 (x).
$$
Using this correspondence,
we would like to understand under what
conditions the zero modes of the original Dirac model (\ref{Ham.gen1}) can
appear,
and what are the corresponding fermion numbers \cite{niemi}.

Let us first make a few remarks on the symmetry properties
of the model (\ref{Ham.gen1}) and some of their implications.
For $m_2 = 0, ~m_1 \neq 0$, the Hamiltonian is invariant with respect
to global phase transformations of the fermion fields,
$
R \rightarrow e^{i \alpha} R, ~~~L \rightarrow e^{i \alpha} L,
$
which amounts to conservation of the total particle number
$$
N = \int dx~ \psi^{\da} (x) \psi (x).
$$
The continuous chiral symmetry
$
R \rightarrow e^{i \gamma} R, ~~~L \rightarrow e^{- i \gamma} L,
$
is broken by the Dirac mass, and the current
$$
J = \int dx~ \psi^{\da} (x) \gamma_5 \psi (x)
$$
is not conserved.
Since $N$ is conserved, the existence of zero modes for a
solitonic shape of $m_1 (x)$ will
lead to the appearance of fractional fermion number \cite{niemi}.

On the other hand, for  $m_1 = 0, ~m_2 \neq 0$, the global phase invariance is
broken by the Cooper-mass
term, and $N$ is not conserved. So, there is no nonzero fermion number in this
case.
However, there is a continuous chiral (or $\gamma_5$) symmetry which is
respected by the Cooper mass term,
giving rise to conservation of the current $J$. Again, if zero modes exist,
a fractional local current
will appear.

This can be understood in two ways. The Cooper mass can be
transformed to a Dirac mass by a particle-hole transformation of one chiral
component of the Dirac
field, e.g. $R \rightarrow R,~L \rightarrow  L^{\da}$. Under this
transformation,
$N \rightarrow J$;
hence the quantized fractional current. The other explanation is physical.
For a solitonic shape of the Cooper mass $m_2 (x)$, the existence of a nonzero
local vacuum
current is a manifestation of the Josephson effect, because when
passing the impurity point the gap function changes its phase (by $\pi$).

Consider now a more general situation, relevant to our discussion of disordered
spin ladders,
when both mass functions $m_1 (x) $ and $m_2 (x)$ have
single-soliton profiles with coinciding centers of the kinks, say, at $x = 0$,
but with arbitrary signs of the corresponding
topological charges and arbitrary amplitudes at $x = \pm \infty$.
Before starting calculations, we can anticipate the characteristic
features  of the solution.
At $m_1 (x), m_2 (x) \neq 0$
none of the above mentioned continuous symmetries survives, i.e.
neither $N$ nor $J$ is conserved.
There are only discrete symmetries present: one is the
$E \rightarrow - E$ symmetry generated by
transformations $R \rightarrow R,~L \rightarrow - L$,
the other being charge conjugation or particle-hole
symmetry:
$
R \rightarrow R^{\da}, ~~L \rightarrow L^{\da}: ~ H \rightarrow H.
$
These symmetries imply that, if localized vacuum states appear
in the solitonic backgrounds of the masses
$m_1 (x)$ and $m_2 (x)$,
those are supposed to be {\it zero} modes, i.e.
localized states exactly
at $E = 0$. Moreover, for massive Dirac fermions ($m_2 = 0$), it is the charge
conjugation symmetry that
implies {\it quantization} of fermion number (see review article
\cite{niemi}):
$\la \hat{N} \ra = n / 2, ~n \in Z$.
For our Hamiltonian (\ref{Ham.gen1}) the first statement still holds: if
localized modes exist, they
should be zero-energy modes. But the second statement is no longer correct:
since $N$ and $J$ are not
conserved, their local expectation values will not be quantized. Instead,
the magnitude of the fermion number
or current will depend on the ratio of the amplitudes $m^0 _1 / m^0 _2$
of the corresponding mass functions. Only in the limiting cases
$m^0 _1 / m^0 _2 \rightarrow \infty$ and
$m^0 _1 / m^0 _2 \rightarrow 0$ will the universal quantized values of the
charge and current be recovered.

Let us now turn to calculations. We consider a single massive Majorana
field described by
the Hamiltonian (\ref{Majorana})
for a fixed index $\sigma$. It can be represented as
\begin{equation}
H_m = \frac{1}{2} \xi^T {\cal H} \xi,
\end{equation}
where
\begin{eqnarray}
{\cal H} &=& \hat{p} v \sigma_1 - m(x) \sigma_2 \nonumber\\
&=&
\left(
\begin{array}{clcr}
      0, && \hat{p}v + i m(x) \\
      \hat{p}v - i m(x), && 0
\end{array}
\right).
\end{eqnarray}
Here $\sigma_1$ and $\sigma_2$ are the Pauli matrices.
Introducing a Majorana 2-spinor
\begin{equation}
u (x) =
\left(
\begin{array}{clcr}
u_R (x) \\
u_L (x)
\end{array}
\right)
\end{equation}
we get a pair of first-order equations:
\begin{equation}
( v \frac{d}{dx} + m(x)) u_R = i E u_L, ~~~
(v \frac{d}{dx} - m(x)) u_L = i E u_R. \label{Maj.eqs}
\end{equation}

Let $m(x)$ have a step-like jump at $x = 0$:
\begin{equation}
m (x) = m_0 sign~x . \label{solitonic-m}
\end{equation}
For $m_0 > 0$ we shall call configuration (\ref{solitonic-m})
a soliton (${\bf s}$); the case
$m_0 < 0$ corresponds to an antisoliton ($\bar{{\bf s}}$).
It immediately follows from Eqs. (\ref{Maj.eqs}) that in the case of a soliton
\begin{equation}
u_{\bf s} (x) = \left(
\begin{array}{clcr}
1\\
0
\end{array}
\right) u_0 (x),
\end{equation}
whereas in the case of an antisoliton
\begin{equation}
u_{\bar{\bf s}} (x) = \left(
\begin{array}{clcr}
0\\
1
\end{array}
\right) u_0 (x).
\end{equation}
Here
\begin{equation}
u_0 (x) = \lambda^{-1/2} _0 \exp ( - |x| / \lambda_0 ), ~~~\lambda_0 = v / m_0,
\end{equation}
is the normalized zero-mode wave function for a bound state of the Majorana
fermion,
appearing
at the discontinuity point of $m(x)$.

Now we have to single out the contribution of zero modes
in the spectral expansion of the Majorana field
operator:
\begin{eqnarray}
\xi_{\bf s} (x) &=& d
\left(
\begin{array}{clcr}
1\\
0
\end{array}
\right) u_0 (x) + \tilde{\xi}_{\mbox{s}} (x), \\
\xi_{\bar{\bf s}} (x) &=& d
\left(
\begin{array}{clcr}
0\\
1
\end{array}
\right) u_0 (x) + \tilde{\xi}_{\bar{\mbox{s}}} (x)
\end{eqnarray}
Here $d$ is the Majorana operator for the zero mode, and
$\tilde{\xi} (x)$ is a contribution of the continuum of scattering states
which,
due to the $E \rightarrow - E$ symmetry, do not affect the
expectation values of fermion number
and current. This part of the Majorana field operator will not be considered
below.

Let us now turn back to the Hamiltonian (\ref{Ham.gen1}).
The fermion number and current operators can be expressed
in terms of the Majorana fields $\xi^{\pm}$ as follows:
\begin{eqnarray}
\hat{N} &=& i ( \xi^+ _R  \xi^- _R  + \xi^+ _L  \xi^- _L )
= i (\xi^+)^T \xi^-,\\   \label{N-Maj}
\hat{N_5} &=& i ( \xi^+ _R  \xi^- _L  + \xi^+ _L  \xi^- _R )
= i (\xi^+)^T \sigma_1 \xi^-, \label{N_5-Maj}
\end{eqnarray}
(remember that $\gamma_5 = \sigma_1$).
There are two qualitatively different cases.

1) Both $m_+$ and $m_-$ have solitonic (or antisolitonic) shapes.

Choose, e.g. $m^0 _+, m^0 _- > 0$. Then
\begin{eqnarray*}
\xi^+ (x) &=& d_+
\left(
\begin{array}{clcr}
1\\
0
\end{array}
\right) u^+ _0 (x), \\
\xi^- (x) &=& d_-
\left(
\begin{array}{clcr}
1\\
0
\end{array}
\right) u^- _0 (x).
\end{eqnarray*}
In this case  one finds that the local current vanishes, while the zero-mode
fermion number is given by
\begin{eqnarray}
N = \la  \hat{N} \ra = \int dx~ \la : \psi^{\da} (x) \psi (x) : \ra
&=& \frac{i}{\cosh \Theta} d_+ d_- \nonumber\\
&=& \frac{1}{\cosh \Theta} ( a^{\dagger} a - \frac{1}{2} ). \label{ferm-num}
\end{eqnarray}
Here we have introduced a local complex fermion operator
$$
a = \frac{d_+ + i d_-}{\sqrt{2}},
$$
while
\begin{equation}
\exp 2 \Theta = m^0 _+ / m^0 _-,
\end{equation}
(This parametrization assumes that both $m^0 _+$ and $ m^0 _-$ are positive,
i.e. $m^0 _1 \geq m^0 _2$.)
The factor $1/\cosh \Theta$ represents the overlap integral between two
zero-mode
wave functions in the
$(+)$ and $(-)$ channels, respectively. One finds that for a pure Dirac mass
$m^0 _2 = 0, ~m^0 _+ = m^0 _-$ and $\Theta = 0$. As a result, the fermion
number
is
quantized:
$$
N = a^{\dagger} a - \frac{1}{2}, ~~\la N \ra = \pm \frac{1}{2}.
$$

2)  $m_+$ has a solitonic shape, while $m_-$ has an antisolitonic shape, or
vice
versa.

We shall choose $m^0 _+ > 0, ~m^0 _- < 0$. Then
\begin{eqnarray*}
\xi^+ (x) &=& d_+
\left(
\begin{array}{clcr}
1\\
0
\end{array}
\right) u^+ _0 (x), \\
\xi^- (x) &=& d_-
\left(
\begin{array}{clcr}
0\\
1
\end{array}
\right) u^- _0 (x).
\end{eqnarray*}
In this case the situation is inverted: the vacuum fermion number vanishes,
while the current does not. The same
calculations lead to
\begin{equation}
J
= \frac{1}{\cosh \Theta} ( a^{\dagger} a - \frac{1}{2} ).
\end{equation}

Let us now turn to the Hamiltonian $H_-$ in Eq. (\ref{TDMHmin}). In this
Hamiltonian
$
m_+ (x) = 3 m (x), ~m_- (x) = - m (x),
$
so that, for $m (x) = m_0~sign~x$ with arbitrary sign of $m_0$,
we are dealing with the case of a nonzero local current (case 2)).
Since $\cosh \Theta = 2 / \sqrt{3}$ in our case, we get:
\begin{equation}
J = \frac{\sqrt{3}}{2} ( a^{\dagger} a - \frac{1}{2} ).
\end{equation}

These simple results can be immediately applied to the smooth parts of
the spin density at the
impurity point. One easily finds that the total spin density
\begin{equation}
S^z _+ (x) = S^z _1 (x) + S^z _2 (x) = \frac{1}{\pi} \p_x \Phi_+ (x) =
: \psi^{\da} (x) \psi (x) :
\end{equation}
while the realtive spin density
\begin{equation}
S^z _- (x) = S^z _1 (x) - S^z _2 (x) = \frac{1}{\pi} \p_x \Phi_- (x) =
: \chi^{\da} (x) \chi (x) :
\end{equation}
So, $\la S^z _+ (x) \ra$ and $\la S^z _- (x) \ra$ coincide with fermion
numbers
in the $(+)$ and $(-)$ channels:
\begin{eqnarray}
\la S^z _+ (x) \ra &=& \frac{1}{2} \sigma u^2 _{m_0} (x), ~~~\sigma = \pm
1), \\
\la S^z _- (x) \ra &=& 0.
\end{eqnarray}
Notice that the fermion number is quantized; hence the localized spin is 1/2
(as can be checked by integrating
$\la S^z _+ (x) \ra$ over $x$).

The spin current operators are defined as
\begin{equation}
{\bf j} (x) = v [ {\bf J}_R - {\bf J}_L ],
\end{equation}
so that we have:
\begin{eqnarray}
\la j^z _+ (x) \ra &=& 0, \nonumber\\
\la j^z _- (x) \ra &=& v ~\sigma~  u_{- m_0} (x) u_{ 3 m_0} (x),
\end{eqnarray}
hence the integrated current
\begin{equation}
\la j^z _-  \ra = \frac{\sqrt{3}}{2} v \sigma .
\end{equation}
Therefore the spin current at the impurity point is not universal and it is
determined
by the
ratio of the
singlet and triplet masses.


\section{Thermodynamic functions for disordered spin ladders}
\label{therm}

\subsection{Free energy}

Since we shall be interested in the behavior of the system in
a magnetic field, we start this section by adding the magnetic
field term to the Hamiltonian (in this paper only a spatially homogeneous
magnetic field will be considered):
\[
-\mu_B H \int \rd x \left[s_1^z(x)+s_2^z(x)\right]
= - \frac{\mu_B}{\sqrt{\pi}}
\int \rd x \partial_x \Phi_+(x)
\]
As follows from (\ref{TDMpsi}),
the magnetic field appears as a chemical potential
(equal to $\mu_BH$)
in the refermionized version (\ref{TDMHDir}) of the $H_+$ part
of the Hamiltonian
\bea
H_D^{m_t}[\psi]&\rightarrow& H_D^{m_t;H}[\psi]=H_D^{m_t}[\psi]
\nonumber\\
&-&\mu_B H\int\rd x :\left[\psi^\dagger_R(x)
\psi^{\phantom{\dagger}}_R(x)+
\psi^\dagger_L(x)\psi^{\phantom{\dagger}}_L(x)\right]:
\label{TFHDirH}
\eea
(the normal ordering in this formula must be taken with respect
to the ground state of the system without  magnetic field).

The magnetic field breaks the spin rotational
invariance
of the problem. Hence it is convenient to work with the
Dirac version
(\ref{TDMHDir}) rather than the Majorana version
(\ref{TDMDir-Maj}),(\ref{TDMMaj}) of the Hamiltonian $H_+$.
We thus rewrite the total Hamiltonian for the
spin ladder in the magnetic field as
\be
H=H_D^{m_t;H}[\psi]+H_M^{m_t}[\zeta^3]+H_M^{m_s}[\zeta^0]\;.
\label{TFHtotH}
\ee

Assume that we know the exact, averaged over the disorder, free energy
for the random mass Dirac fermions
($H_D^{m;H}[\psi]$) as a function of
the magnetic field $H$ and temperature $T$.
This free energy, which we denote by $F_D^m(T,H)$,
will be actually calculated  in what follows.
Clearly the free energy of random
mass Majorana fermions $F_M^m(T)$ can be
inferred from the above function (notice that there is no
magnetic field term for a single Majorana fermion).
Since $H_D^{m;H=0}$ decomposes into two Hamiltonians $H_M^m$
for independent Majorana fields
and the free energy is an extensive variable, we simply have
\[
F^m_M(T)=\frac{1}{2}F_D^m(T,0)\;.
\]
The total free energy of the system (\ref{TFHtotH}) is therefore
\be
F(T,H)=F_D^{m_t}(T,H)+\frac{1}{2}F_D^{m_t}(T,0)+\frac{1}{2}
F_D^{m_s}(T,0)\;.
\label{TFFtot}
\ee

In the following we shall focus on the function $F_D^m(T,H)$.
Since the fermions are noninteracting, the
regularized free energy can be
written as
\bea
&~&~\Delta F_D^m(T,H)=F_D^m(T,H)-F_D^m(0,0)=\label{TFFDir}\\
&~&~-T\int\limits_{-\infty}^{0}\rd\epsilon
\rho_D(\epsilon) \ln\left[1+
{\rm e}^{\beta\left(\epsilon-\mu_BH\right)}
\right]
-T\int\limits_{0}^{\infty}\rd\epsilon
\rho_D(\epsilon) \ln\left[1+
{\rm e}^{\beta\left(\mu_BH-\epsilon\right)}
\right]
\;,\nonumber
\eea
where $\beta$ is the inverse temperature, and
$\rho_D(\epsilon)$ is the
density of states for the Dirac fermions (\ref{TDMHDir}),
averaged over the disorder\cite{RG}.

In fact, $\rho_D(\epsilon)$ is a single particle density of
states for the quantum mechanical problem of a Dirac particle
with a random mass\cite{mcK}. The wave--function of the latter, for
which we keep the same notation as for the field operator
\[
\psi(x)=\left[
\begin{array}{c}
\psi_R(x)\\
\psi_L(x)
\end{array}
\right]\;,
\]
satisfies the Dirac equation of the form
\be
\left[-iv_t{\sigma}_3\partial_x+m(x){\sigma}_2\right]
\psi=\epsilon\psi\;,
\label{TFDireq}
\ee
[there is
no chemical potential term in Eq. (\ref{TFDireq}), for it has been explicitly
separated in (\ref{TFFDir})].
It is convenient to make the chiral rotation (\ref{chir.rot})
\[
\psi_{R,L}(x)=\frac{1}{\sqrt{2}}\left[
v(x)\pm u(x)\right]\;,
\]
The spinor component $u(x)$ then satisfies the Schr\"odinger--type
equation
\be
\left[-\partial_x^2+\bar{m}^2(x)-\bar{m}'(x)\right]u=Eu\;,
\label{TFWitt}
\ee
where $\bar{m}(x)=m(x)/v_t$ and $E=\epsilon^2/v_t^2$.
The spinor component $v(x)$ satisfies Eq. (\ref{TFWitt}) with
$m(x)$ replaced by $-m(x)$.

The equation (\ref{TFWitt}) is known as Witten's toy model for
supersymmetric quantum mechanics \cite{Witt}.
To our great advantage, this problem with a telegraph signal $m(x)$
has recently been solved exactly
by Comtet, Desbois, and Monthus (CDM) \cite{CDM}.
In particular,
CDM calculated the disorder-averaged
 integrated density of states $N(E)$ for
the problem (\ref{TFWitt}).
Comparing Eq.(\ref{TFWitt}) and Eq.(\ref{TFDireq}),
one easily finds that the density of states for the Dirac particle
we are interested in is related to the integrated density of states
of Ref.\cite{CDM} by
\be
\rho_D(\epsilon)=\frac{2\epsilon}{v_t^2}
N'
\left(
\frac{\epsilon^2}{v_t^2}\right)\;.
\label{TFDOSrel}
\ee

Thus we have, in principle, an exact solution for
the free energy of the disordered spin ladder.
The analytical expression for the function $N(E)$ found by
CDM is, however, quite complicated.
Therefore,
instead of reproducing this expression here
(an interested reader is referred to the original paper \cite{CDM}),
we shall consider particular limiting cases.

\subsection{Low energy thermodynamics}
\label{TLE}

In this section we consider the thermodynamic functions of the
disordered spin ladder at the lowest energy scale, $\epsilon\ll
\epsilon_0$, with $\epsilon_0$ to be determined later.

According to the CDM solution and Eq.(\ref{TFDOSrel}),
in the $\epsilon\rightarrow 0$ limit,
the density of states takes the form
(see also discussion in the next section)
\be
\rho_D(\epsilon)\simeq
\frac{2\sigma_0}{\epsilon\ln^3(1/\epsilon)}\;,
\label{TLEDOS}
\ee
where
\be
\sigma_0=\frac{m^2}{v_t^2n_0}\;.
\label{TLEsigma}
\ee
The expression (\ref{TLEDOS}) is given in the leading
logarithmic approximation.

Using (\ref{TFFDir}) together with the particle--hole
symmetry of the problem, implying that
$\rho_D(\epsilon)=\rho_D(-\epsilon)$,
the magnetic moment of the system,
$M(T,H)$, induced by the external
magnetic field is found to be
\be
M(T,H)=\mu_B\int\limits_0^\infty \rd\epsilon
\rho_D(\epsilon)
\left[f(\epsilon-\mu_BH)-f(\epsilon+\mu_BH)\right]\;,
\label{TLEM}
\ee
where $f(\epsilon)=\left[{\rm e}^{\beta\epsilon}+1\right]^{-1}$
is the Fermi function.

The linear magnetic susceptibility therefore is
\be
\chi_l(T)=\frac{\mu_B^2}{2T}
\int\limits_0^\infty \rd\epsilon
\rho_D(\epsilon) \cosh^{-2}
\left[\frac{\epsilon}{2T}\right]\;.
\label{TLEchil}
\ee
As the temperature is lowered, the linear
susceptibility diverges as
\be
\chi_l(T)\simeq\frac{\mu_B^2\sigma_0}{2T\ln^2(1/T)}\;.
\label{TLEchilT}
\ee
This can be interpreted as a
Curie like behavior with a vanishing
Curie constant,
$C(T)\simeq \mu_B^2\sigma/2\ln^2(1/T)$.

On the other hand, the zero--temperature magnetic moment
is simply proportional to the integrated density of
states
\be
M(0,H)=\mu_BN\left[\frac{\mu_B^2H^2}{v_t^2}\right]\;.
\label{TLEMH}
\ee
For a weak magnetic field
\be
M(0,H)\simeq
\frac{\mu_B\sigma_0}{\ln^2(1/H)}\;,
\label{TLEMsmallH}
\ee
leading to a nonlinear susceptibility
\be
\chi_n(H)=\frac{M(0,H)}{H}\simeq\frac{\mu_B\sigma_0}{H\ln^2(1/H)}\;,
\label{TLEchin}
\ee
which diverges in the same way as the
linear susceptibility (\ref{TLEchilT}) does.
This indicates that the magnetic field scales as the temperature,
as it should be for noninteracting particles.
The differential susceptibility, however,
directly measures the density of states
\be
\chi_d(H)=\frac{\partial
M(0,H)}{\partial H}=\mu_B^2\rho_D(\mu_BH)
\simeq\frac{2\mu_B\sigma_0}{H\ln^3(1/H)}\;.
\label{TLEchid}
\ee
Interestingly, the low--temperature correction to the
magnetic moment in a finite field is also  a
highly singular function of the field
\be
M(T,H)-M(0,H)\simeq-
\frac{\pi^2\sigma_0 T^2}{\mu_B H^2 \ln^3(1/H)}\;.
\label{TLEMHT}
\ee

The zero--field free energy reads
\be
\Delta F_D(T,0)=-2T\int\limits_0^\infty\rd\epsilon
\rho_D(\epsilon)\ln\left(1+{\rm e}^{-\beta \epsilon}\right)\;.
\label{TLEFT}
\ee
The low-temperature free energy correction is therefore
\be
\Delta F_D(T,0)\simeq-\frac{2\ln 2\sigma_0 T}{\ln^2(1/T)}\;.
\label{TLEFsmallT}
\ee
An unusual behavior of the specific heat follows:
\be
C_V(T)\simeq\frac{8\ln 2\sigma_0 \eta}{\ln^3(1/T)}\;,
\label{TLECV}
\ee
where the parameter
\[
\eta=\frac{3v_s^2+v_t^2}{4v_s^2}
\]
is a function of the ratio of the velocities in the singlet
and the triplet sectors.

The low-temperature entropy vanishes as
\be
S(T)\simeq\frac{4\ln 2 \sigma_0 \eta}{\ln^2(1/T)}\;,
\label{TLES}
\ee
indicating a non--degenerate ground state.
However, the entropy vanishes very slowly
with  temperature reflecting the presence
of `quasi--free moments' in the system
(see also the next Section).

According to (\ref{TLECV}), the specific heat
coefficient, $C_V(T)/T$, diverges as $T\to 0$.
Yet the specific heat coefficient is less
divergent than the linear magnetic susceptibility
(\ref{TLEchilT}). Hence a very large Wilson ratio
\be
R_W(T)=\frac{T\chi_l(T)}{C_V(T)}
\simeq\frac{\mu_B^2}{16\ln 2 \eta}\ln (1/T)\;.
\label{TLEWR}
\ee
We notice that if one instead associates the Wilson ratio
with the differential magnetic susceptibility (\ref{TLEchid}),
then this modified Wilson ratio becomes a constant,
only depending on the ratio of the velocities:
\be
\tilde{R}_W(\bar{T})=\frac{\bar{T}\chi_d(\bar{T})/\mu_B}{
C_V(\bar{T})}=\frac{\mu_B^2}{\ln 2}
\frac{v_s^2}{3v_s^2+v_t^2}
\label{TLEWRbis}
\ee

\subsection{Intermediate regime}
\label{TIR}

The singularity in the density of states of the form
(\ref{TLEDOS}) has been obtained by Dyson
back in 1953 \cite{Dyson} for a model of disordered
harmonic chain.
In the electronic spectrum at the centre of the Brillouin
zone such a singularity was identified by
Weissman and Cohan \cite{WC} for the case
of a non--diagonal disorder (random hopping model).
The latter model is in fact directly related to
the random mass Dirac problem through the notion
of zero--modes (see Section \ref{ZM} and also below).
The Dyson singularity in the density of states
persists the case of a
half--filled electron band
with random backscattering, as shown
by A.A. Gogolin and Mel'nikov \cite{AAG}
who also obtained the low-temperature asymptotics for the
magnetic susceptibility (\ref{TLEchilT})
to explain experimental data on
$TCNQ$ salts.
A similar low-temperature magnetic susceptibility
has been predicted by Fabrizio and Melin \cite{FM}
for inorganic
spin--Peierls compounds $Cu Ge O_3$.
It must be pointed out that the spin--Peierls systems,
sharing with the spin ladders the property
of having the spin gap,  behave in quite
a similar way under doping \cite{SPrev}.

The low energy behavior (\ref{TLEDOS}) is characteristic
for various particle--hole symmetric models of disorder
and most probably represents a universality class.
For the random mass Dirac problem such a behavior was found
by Ovchinnikov and Erikhman \cite{OE}
assuming a Gaussian white noise distribution
of the mass variable $m(x)$.
The advantage of the CDM solution for a telegraph signal mass,
which incorporates the Gaussian distribution as a
particular case,
is that it keeps track of the high--energy properties
extrapolating between the universal
low--energy regime ($\epsilon\ll\epsilon_0$)
effectively described by the Gaussian distribution,
and
the high--energy regime ($\epsilon\gg m_0$) of almost free massless particles.
This enables us to describe the intermediate regime
($\epsilon_0\ll\epsilon< m_0$).
The latter only exists for low impurity concentration.
Indeed, CDM found that for $n_0\to 0$ the integrated
density of states, after an initial increase at low
energies, saturates to the value
\be
N(\epsilon)\simeq \frac{n_0}{2} \;\;\;{\rm at}\;\;\;
\epsilon_0\ll\epsilon< m_0\;.
\label{TIRNsat}
\ee
Consequently, the density of states $\rho_D(\epsilon)$ almost
vanishes in this region.
 From (\ref{TLEDOS}) and (\ref{TIRNsat}) we can roughly
estimate the crossover energy as
\be
\epsilon_0\sim m_0 \exp
\left(-\frac{\sqrt{2}m_0}{vn_0}\right)\;.
\label{TIRcrossover}
\ee

Let us now consider the disordered spin ladder at temperatures
$\epsilon_0\ll T<m$. From Eq.(\ref{TLEchil})
for the magnetic susceptibility we obtain
\be
\chi_l(T)\simeq\frac{\mu_B^2 n_0}{4T}\;.
\label{TIRchi}
\ee
This is exactly equal to a magnetic susceptibility of
free spins S = 1/2 with concentration $n_0$.
This is in agreement with our discussion of
zero modes in Section \ref{ZM}.
Eqs (\ref{TLEchilT}) and (\ref{TIRchi}) mean that the
Curie constant, being almost temperature independent for
$\epsilon_0\ll T<m_0$, is quenched in the region of
temperatures smaller than $\epsilon_0$.
This behavior is different from the one found by
Sigrist and Furusaki \cite{SF} who, in particular,
predicted a finite Curie constant at $T\to 0$. We will
compare in Section \ref{comp} the recent QMC results\cite{miya} with theoretical
predictions which seems to show that our prediction is confirmed.

It is instructive to consider the free energy correction
(\ref{TLEFT}) at $\epsilon_0\ll T<m$,
from which it follows that the entropy
\be
S(T)\simeq 2\ln 2 n_0 \;.
\label{TIRS}
\ee
Bearing in mind that a local Majorana fermion has a residual
entropy $\ln\sqrt{2}$, we conclude that the expression
(\ref{TIRS}) indicates the presence of four local
Majorana fermions at each impurity location.
Clearly three of these local Majorana fermions originate
from the bulk triplet mode, while the remaining one is due to the
singlet mode.
An effective Hamiltonian for the local Majorana fermions
(which we denote as $\xi^i_a$)
can be written on purely phenomenological grounds.
Indeed, the effective Hamiltonian, that respects all symmetries
of (\ref{IHHtotM}), takes into account the fact that the
magnetic field couples to the $a=1,2$ components of the
triplet mode, and finally preserves the quadratic nature of the
problem, must be of the form
\be
H_{eff}=\sum\limits_{i,j}\left[
\sum\limits_{a=1}^3\tau_{i,j}^t d^i_a d^j_a+
\tau_{i,j}^s d^i_0 d^j_0\right]
-\mu_BH\sum\limits_i d^i_1 d^i_2\;,
\label{TIRHeff}
\ee
where $\tau^{t(s)}_{i,j}$ are the random hopping integrals
related to the overlaps of the zero modes in the triplet
(singlet) sector, while $d^i$ are the individual Majorana
zero--modes operators studied in Section \ref{ZM}.
The expression (\ref{TIRHeff})
clarifies the relation of our random mass problem
with the non--diagonal disorder problem of Ref.\cite{WC}
and hence with the original Dyson model \cite{Dyson}.

\subsection{Comparison of theory with quantum Monte
Carlo simulations}
\label{comp}

We have derived  above formulas for various thermodynamic quantities
at low temperatures. It would be very important to check these
formulas directly by experiments. Unfortunately, these experiments are very
difficult
 to perform: one reason being the interference of various
factors (lattice, other impurities, etc); another reason being the smallness
of the logarithmic corrections.  On the other hand, there has been
just performed a set of very nice QMC simulations on doping effect in
two-leg ladders\cite{miya}. Those authors could simulate  very large
systems (up to 2000 sites) and get down to very low temperatures ($T=0.005J$).
In particular, they could carry out  random average over   different impurity
configurations (up to 20 realizations) which is a crucial factor
in comparison with analytic theory (which, of course, assumes random
distribution).

In Fig.1 we  replotted their numerical data on uniform magnetic susceptibility
(Fig. 2 in their original paper) vs $1/(\ln T)^2$ (The temperature $T$
is measured in units of $J$). The fitting formula is
$ \chi T\; =\; c\; [\;a\; +\; b/(\ln T))^2 ]$, where $c$ is the
doping concentration, while the  parameters
$a = 0.185(3),\; 0.145(2),\; 0.126(1),$
and  $b = 0.43(3),\; 0.29(1),\; 0.23(1)$ for
$c = 0.01,\; 0.05,\; 0.1 $, respectively.
It is  important to note that according to RSRG considerations\cite{SF}
$a$ should be 1/12 at very low temperatures, and 1/4 at intermediate
ones (as indicated by arrows and dotted lines in Fig.1),
 whereas $b$ should be zero. On the contrary, according
to our analysis, eq.  (\ref{TLEchilT}) $a$ should be zero, while $b$
should be 1/2. The numerical results do  clearly show the presence of the
logarithmic term, but the effective Curie constant does not vanish
entirely as $ T \rightarrow 0$: There is a finite intersection $a \neq 0$,
and the slope $b$ is less than 1/2 as expected.
As mentioned in the previous subsection, the asymptotic logarithmic
behavior should be valid for $T \le \epsilon_0$, where $\epsilon_0$ is
the low energy scale
in the problem (roughly speaking, the soliton band width).
Therefore, one should anticipate good agreement only at very low temperatures
( when $1/(\ln T)^2 \ll 1$). It is
quite interesting to notice that the linear fitting is better (over broader
temperature range) for higher concentrations,
where $\epsilon_0$ is larger.

Our tentative
interpretation for the absence of full agreement with the theoretical
prediction
 is due to the fact that the random sampling in \cite{miya}
is still not big enough to fully demonstrate the anticipated behavior.
	To explain this point, let us recall the basic physical picture of the
zero energy states in the Dirac model with random mass. Some of these
states are genuinely localized, while the others are only "quasilocalized".
The first category of states is "typical", while in taking random average
the second
category of states does contribute in a substantial way.
These features  show up clearly in the spin-spin
correlations calculated using the Liouville quantum mechanics\cite{Liouville},
 the Berezinskii diagram technique\cite{Berezinskii}, and the
supersymmetric method
\cite{supersymmetry}. The typical
configurations for the spin correlation functions are exponentially decaying,
whereas the average behavior has a  power-law decay. The difference between
the "typical" and "average" configurations is the nontrivial piece of
physics involved in randomly doped gapped spin systems (as well as  some
other random systems).  The fact that the density of states shows the Dyson
singularity and other thermodynamic quantities show logarithmic
singularities are
all due to the same reason.  It is quite remarkable that the signature of
this behavior has shown up in the QMC simulations.  As clear from the
above explanation, the logarithmic singularity will show up as  "full-fledged"
only if the random sampling is really big. Otherwise, we will  still see
some constant term $a$ as "remnant" feature of the dominance of exponentially
decaying states. Hopefully, with the further improvement of the numerical
techniques, this prediction can be checked more precisely. Namely, when the
sampling
 becomes bigger and bigger, the intersection with the vertical axis
(the remaining part of the Curie constant) should gradually vanish.
To the best of our knowledge, there is no  explanation for
this type of logarithmic singularities other than the one described above.
Therefore the presence of this term per se in the numerical simulations
is already significant.

\section{Staggered magnetization near the defects}
\label{SS}

 In fact, in the continuous Majorana model there are two vacuum
averages: the staggered magnetization and the smooth magnetization
are both nonzero in the vicinity of the point where $m(x)$ changes
sign. Unfortunately, we  have not been able to calculate the staggered
magnetization for the model with sign-changing $m(x)$; we have done it
only in the
model with a sharp edge (that at the end of a broken
chain). Nevertheless since this solution shows the presence of zero
modes we think that it gives a qualitatively correct description of
the staggered magnetization. 

The
calculation is based on two facts. 
The first one is that the order and disorder parameters
in the Ising model are expressed
in terms of fermion creation and annihilation operators $R(\theta)$,
$R^+(\theta)$ as follows ($T
> T_c$) \cite{Sato}:
\[
\mu(\tau, x) = :\exp[\frac{1}{2}\rho_F(\tau, x)]: \:
\sigma(\tau, x) = :\psi_0(\tau,x)\exp[\frac{1}{2}\rho_F(\tau, x)]:
\]
\[
\rho_F(\tau, x) = - i\int_{-\infty}^{\infty} d\theta_1 d\theta_2
\tanh[(\theta_1 - \theta_2)/2]\exp[(\theta_1 + \theta_2)/2]
\]
\bea
\times\exp[ -
m\tau(\cosh\theta_1 + \cosh\theta_2) - i mx(\sinh\theta_1 +
\sinh\theta_2)]R(\theta_1)R(\theta_2) \nonumber\\
+ \mbox{terms with } R^+
\eea
\bea
\psi_0(\tau,x) = \int_{-\infty}^{\infty} d\theta
\{e^{\theta/2}\exp[- m\tau\cosh\theta - i mx \sinh\theta]R(\theta)\nonumber\\
+ \mbox{term with } R^+\}
\eea
These fermion operators satisfy the standard anticommutation
relations:
\[
[R(\theta), R^+(\theta')]_+ = \delta(\theta - \theta')
\]
which,  in the case of the Ising model represent the simplest
realization of the Zamolodchikov-Faddeev algebra.

 Since the operators are normally ordered and $<0|R^+(\theta) = 0$
in $< 0|\mu $ and $<0|\sigma$ we can omit all terms with $R^+$:
\[
<0|\mu(\tau, x) = < 0|\exp\{\frac{- i}{2}\int_{-\infty}^{\infty}d\theta_1
d\theta_2
\tanh(\theta_{12}/2)\exp[(\theta_1 + \theta_2)/2]\exp[ -
m\tau(\cosh\theta_1 + \cosh\theta_2)
\]
\bea
- i mx(\sinh\theta_1 +
\sinh\theta_2)]R(\theta_1)R(\theta_2)\} \label{mu}
\eea
\[
< 0|\sigma(\tau, x) = \la 0|\int_{-\infty}^{\infty} d\theta
[ e^{\theta/2}\exp[- m\tau\cosh\theta - i mx
\sinh\theta]R(\theta)
\]
\bea
\times\exp\{\frac{- i}{2}\int_{-\infty}^{\infty} d\theta_1 d\theta_2
\tanh(\theta_{12}/2)\exp[(\theta_1 + \theta_2)/2]\exp[ -
m\tau(\cosh\theta_1 + \cosh\theta_2) \nonumber\\
- i mx(\sinh\theta_1 +
\sinh\theta_2)]R(\theta_1)R(\theta_2)\} \label{sigma}
\eea

  The second fact is that in the
approach suggested by Ghoshal and Zamolodchikov \cite{GZ}
time and space coordinates are interchanged and  the  boundary
is thought about as an asymptotic state at $t \rightarrow
\infty$. This out-state is denoted  as  $|B>$.  Each integrable
model has its own $|B>$-vector. For the Ising model with
free boundary conditions
can be represented by the  state vector is given by
\bea
|B> = [1 + R^+(0)]\exp\{ -
\frac{ i}{2}\int_{-\infty}^{\infty} d\theta\coth(\theta/2)R^+(-
\theta)R^+( \theta)\}|0> \label{state}
\eea
Notice that it contains a fermionic creation operator with zero
rapidity; this operator corresponds to the Majorana zero mode - a
boundary bound state. This mode is the {\bf zero energy}
described in the previous sections.  Since in the
Ghoshal-Zamolodchikov formalism space and time are interchanged,
$R^+(0)$ formally creates a state with zero momentum.

 Let us calculate a vacuum average of $\mu$ at point $X$:
\bea
&~&~<\mu(t, X)> = <\mu(\tau = X, x = t)|B>\nonumber\\
&~&~< 0|\exp[ - i\int_{-\infty}^{\infty} d\theta_1 d\theta_2
A(\theta_1,\theta_2)R(\theta_1)R(\theta_2)]\exp[ -
\frac{i}{2}\int_{-\infty}^{\infty} d\theta\coth(\theta/2)R^+(-
\theta)R^+( \theta)\}|0>\nonumber\\
&~&~= \exp[ - \frac{1}{2}\int_{-\infty}^{\infty} d\theta A(\theta, -
\theta)\coth(\theta/2)]
\eea
where $A(\theta_1,\theta_2)$ is defined in (\ref{mu}).
Since the exponents commute on  constant, we can use the formula
\be
e^A e^B = e^B e^A e^{[A, B]}
\ee
and  obtain the following result:
\bea
<\mu(X)> = \exp( - \frac{1}{8}\int_{0}^{\infty}d\theta(1 +
1/\cosh\theta)e^{- 2mX\cosh\theta}) \nonumber\\
=  \exp\{- \frac{1}{8}[K_0(2mX) + K_{-1}(2mX)]\}\\
<\sigma(X)> = \exp( - m|X|)\exp\{- \frac{1}{8}[K_0(2mX) +
K_{-1}(2mX)]\}
\eea
that is
\bea
n(X) = <\mu_1(X)\sigma_2(X)\sigma_3(X)\mu_0(X)> \nonumber\\
= \exp( - m|X|)\exp\{-
\frac{1}{8}[3K_0(2mX) + K_0(6mX)
3K_{-1}(2mX) + K_{-1}(6mX)]\} \label{stag}
\eea
This vacuum average behaves as $X^{-1/2}$ at $X << m^{-1}$ and decays
exponentially at $X >> m^{-1}$.

\section{Conclusions and Discussion}
\label{CD}

In this paper we have shown that doped spin-1/2 ladder
systems are described by the random mass Majorana (Dirac)
fermion model.
On the basis of this model,
we have calculated the thermodynamic functions
for the spin ladders.
In particular, we predict $1/T\ln^2 T$ low--temperature asymptotics
for the linear magnetic susceptibility.
This behavior is quite different from the
simple Curie law.
As discussed in Section\ref{comp}, there is already good
evidence of this behavior in numerical simulations.
Of course, we hope that more precise experimental
measurements would be able to distinguish these
nontrivial disorder effects from the
contributions of uncorrelated 1/2 spins induced by impurities.
 We would like also to point out
that the recent neutron data\cite{Zudop1} have shown the existence
of the gap states, while the magnitude of the gap itself does not change
with doping. This is certainly consistent with our theoretical
results.

We did not attempt to discuss more complicated questions
related to the behavior of
the correlation functions in such  disordered
systems.
We only note here that the divergency of the density
of states (and of the localization length)
in the middle of the gap makes these
systems different from standard one-dimensional
disordered systems giving rise to a non-trivial
critical regime at low energies \cite{AAG,RG}.
In fact, in the recent months, there has been quite an
impressive progress in the understanding of
the correlation functions.
For instance, some important insight
into the zero energy properties of the
random mass Dirac model was provided by
 mapping of this problem onto
a Liouville quantum mechanics \cite{Liouville}.
An interplay between the critical regime at low
energies and the standard localization regime
was explored by Beresinskii's diagram technique \cite{Berezinskii}
and by the supersymmetry method \cite{supersymmetry}.

However, the influence of the disorder on the
staggered magnetic susceptibility in spin ladders has been poorly
studied so far.
This quantity is important from
the experimental point of view.
It is vital for the understanding of the apparent promotion
of the antiferromagnetic
ordering upon doping, which was experimentally
observed in both the spin ladder and the
spin--Peiels systems \cite{Zndop,SPrev}.
It is our opinion that future theories
of the antiferromagnetic transition
in these systems will be based on the
mapping onto the random mass fermion model,
presented in this paper, in conjunction
with the theoretical progress in dealing with
such fermionic models achieved in
Refs \cite{Liouville,Berezinskii,supersymmetry}.

\section{Acknowledgments}

We are thankful to M. Fabrizio, D. Abraham  and J. Gehring for illuminating
discussions and useful remarks.
We also  sincerely thank the authors of \cite{miya} for kindly
sending us their QMC data on uniform magnetic susceptibility and Dr.
 S.J. Qin for the  help in making the plot.
 A.O.G. was supported by the
EPSRC of the U.K. and part of this work has been done during
his stay at the ICTP, Trieste, Italy. A.A.N. acknowledges the
support from ICTP, Trieste, Italy and A. M. T. thanks ICTP for kind
hospitality. 


\newpage

\figure{  The QMC simulation data of the doping effect on uniform magnetic
susceptibility in spin ladders of Ref.\cite{miya} are replotted as a
function of $1/(\ln T)^2$ in comparison with Eq.(\ref{TLEchilT}).
The temperature $T$ is measured in units of $J$.
The fitting formula is $\chi T\;= c\;[\; a\; +\; b/(\ln T)^2] $,
where $a = 0.185(3),\; 0.145(2),\; 0.126(1),$
and  $b = 0.43(3),\; 0.29(1),\; 0.23(1)$ for doping concentrations
$c = 0.01,\; 0.05,\; 0.1 $, respectively.
The dotted lines are expectations for uncorrelated
free 1/2 spins induced by impurities, while the arrows on the left side
indicate
 the renormalized values, anticipated from the renormalization group
analysis\cite{SF}}.


\begin{thebibliography}{99}
\bibitem{fuk2}
The quantization of charge in the problem under consideration was
first pointed out in \cite{Fuk}. However, ignoring one of the terms
in their hamiltonian (1) breaks the  spin $SU(2)$ symmetry.
\end{thebibliography}
\end{document}